\documentclass[times,5p]{elsarticle}

\usepackage{amsfonts}
\usepackage{amsmath}
\usepackage{subfigure}
\usepackage{algorithm}
\usepackage{algpseudocode}
\usepackage{hyperref}
\usepackage{graphicx}

\newcommand{\del}{\partial}

\newcommand{\Idx}{\mathcal{I}}
\newcommand{\Jdx}{\mathcal{J}}
\newcommand{\ctI}{\mathcal{T}_\Idx}
\newcommand{\ctJ}{\mathcal{T}_\Jdx}
\newcommand{\ctIJ}{\mathcal{T}_{\Idx\times\Jdx}}
\newcommand{\bbbr}{\mathbb{R}}
\newcommand{\bbbc}{\mathbb{C}}

\newcommand{\sons}{\mathop{\operatorname{sons}}\nolimits}
\newcommand{\supp}{\mathop{\operatorname{supp}}\nolimits}
\newcommand{\diam}{\mathop{\operatorname{diam}}\nolimits}
\newcommand{\dist}{\mathop{\operatorname{dist}}\nolimits}

\newtheorem{theorem}{Theorem}
\newtheorem{remark}[theorem]{Remark}

\algnewcommand\AlgAnd{\textbf{and }}

\begin{document}

\begin{frontmatter}

\title{Approximation of boundary element matrices using GPGPUs and
       nested cross approximation}
\author[CAU]{Steffen B\"orm}
\corref{cor}
\ead{boerm@email.uni-kiel.de}
\author[CAU]{Sven Christophersen}
\ead{svc@informatik.uni-kiel.de}
\fntext[DFG]{Part of this work was funded by the
   Deutsche Forschungsgemeinschaft in the context of project
   BO~3289/2-1.}
\address[CAU]{Department of Computer Science,
  {Christian-Albrechts-Universit\"at} zu Kiel, 24118 Kiel, Germany}
\cortext[cor]{Corresponding author.}

\begin{abstract}
\noindent
The efficiency of boundary element methods using a Galerkin discretization
depends crucially on the time required for setting up the
stiffness matrix.
The far-field part of the matrix can be approximated by compression
schemes like the fast multipole method or $\mathcal{H}$-matrix
techniques.
The near-field part is typically approximated by special quadrature
rules like the Sauter-Schwab technique that can handle the singular
integrals appearing in the diagonal and near-diagonal matrix elements.

Since computing one element of the matrix requires only a small
amount of data but a fairly large number of operations, we propose
to use general-purpose graphics processing units (GPGPUs) to handle
vectorizable portions of the computation:
near-field computations are ideally suited for vectorization and
can therefore be handled very well by GPGPUs.
Modern far-field compression schemes can be split into a small adaptive
portion that exhibits divergent control flows, and should therefore be
handled by the CPU, and a vectorizable portion that can again be sent
to GPGPUs.

We propose a hybrid algorithm that splits the computation into tasks
for CPUs and GPGPUs.
The method presented in this article is able to reduce the setup time
of boundary integral operators by a significant factor of 19--30 for 
both the Laplace and the Helmholtz equation in 3D when using 
two consumer GPGPUs compared to a quad-core CPU.
\end{abstract}


\end{frontmatter}


\section{Introduction}

Compared to the popular finite element methods, boundary element
methods offer advantages like improved accuracy and the possibility to
handle infinite domains, but these advantages come at the price of a
stiffness matrix that is no longer sparse and requires us to evaluate
singular integrals.

During the last decades, both issues have been the focus of significant
research activities:
approximation schemes like the fast multipole method
\cite{RO85,GRRO87,AN92,GIRO02,BIYIZO04,OFSTWE06},
panel clustering \cite{HANO89,SA99},
or hierarchical matrices
\cite{HA99,HAKH00,BERJ01,GRHA02,BOGR04} replace the stiffness matrix
by a data-sparse approximation of arbitrary accuracy that can be
computed in almost linear complexity.
State-of-the-art compression schemes relying on $\mathcal{H}^2$-matrices
\cite{HAKHSA00,BOHA02,BO10,BEVE12,BOCH14} can handle matrices with
millions of degrees of freedom in minutes on modern computers.

Graphics processors, originally developed to speed up
CAD applications and computer games, have in the past years been
generalized to handle far more general tasks.
\emph{General-purpose graphics processing units} (GPGPUs) offer
computational resources far outstripping any CPU, e.g., thousands
of arithmetic-logical units combined with high-bandwidth storage,
but at the price of severe restrictions, e.g., only a limited
amount of directly accessible memory and high sensitivity to divergent
control flows inherited from their vector-oriented architecture.

Due to the computational power of GPGPUs, there have been attempts
to utilize the computational power of GPGPUs to reduce setup times for 
discretized boundary integral operators 
\cite{Wang15,LABAKI11,HAMS13,TAHA09}.

Modern compression techniques, e.g., hierarchical matrices
($\mathcal{H}$-matrices) or $\mathcal{H}^2$-matrices rely on
adaptivity to obtain close to optimal compression rates.
An example is the adaptive cross approximation (ACA) method
\cite{BERJ01,BEVE12} that combines a heuristic pivoting
strategy with a rank-revealing LR factorization.

Unfortunately, adaptive algorithms are generally not well-suited
for GPGPUs:
graphics processors are typically designed using a \emph{single
instruction, multiple data} (SIMD) architecture, i.e., groups of
processing elements have to execute \emph{the exact same instruction}
in each clock cycle.
Modern GPGPUs extend this to a \emph{single instruction,
multiple threads} (SIMT) approach that essentially allows some
of the processing elements in a group to do nothing while the
others execute one instruction.
Adaptive algorithms typically iterate until a certain accuracy
has been reached, i.e., the number of iterations depends on the
input data.
In an SIMD/SIMT architecture, this means that the run-time is
determined by the slowest member of a group.

In order to avoid this problem, we propose a hybrid approach:
adaptive parts of the algorithm are performed by CPUs, while
non-adaptive parts like geometry transformations and numerical
integration (qua\-dra\-ture) are handled by GPGPUs.
The recently developed GCA compression scheme based on Green's
representation formula, quadrature, and nested cross approximation
\cite{BOGO12,BOCH14} allows us to significantly reduce the adaptive
parts of the algorithm and to take full advanage of the GPGPUs.

GCA can be split into an adaptive preparation phase that chooses
suitable pivot elements for clusters and computes a number of
small matrices, and a non-adaptive setup phase that computes
matrix entries for the chosen pivot elements by qua\-dra\-ture.

We use the popular Sauter-Schwab quadrature technique
\cite{SCWE92a,SA92,ERSA98,SASC11}, since it relies mostly on
arithmetic operations (addition, subtraction, multiplication,
reciprocal square root) that modern GPGPUs can handle very
efficiently.
Another advantage of this quadrature technique is that each
integral only requires a small amount of data, essentially
only the coordinates of the vertices of two triangles, so there
is a good chance that the memory bandwidth will not limit the
performance of the central quadrature loop.

The Sauter-Schwab method distinguishes four cases:
two triangles can be identical, they can share a common edge
or a common vertex, or they can be completely disjoint.
For each of these four cases, a number of quadrature points
and weights are defined, with the ``highly singular'' first
two cases requiring significantly more points than the last
two cases.
Heterogeneous systems are a good match for these rules:
the main processor classifies the integrals according to the four
categories and sorts them into work packages for the GPGPUs.
Since all elements of a work package correspond to \emph{exactly}
the same quadrature rule, they are ideally suited for vector
processors, and get sent to GPGPUs.
Once all integrals in a work package have been handled, the results
are returned to the main processor to be added to the final stiffness
matrix.
Even with comparably inexpensive mainstream GPGPUs, this approach can
drastically reduce the runtime for singular and near-singular integrals.


\section{Sequential algorithms}

We consider the construction of an $\mathcal{H}^2$-matrix approximation
of the stiffness matrix of a Galerkin BEM discretization.

\subsection{Galerkin discretization}

Given finite element bases $(\varphi_i)_{i\in\Idx}$ and $(\psi_j)_{j\in\Jdx}$
for the test and trial spaces,
a kernel function $g:\bbbr^d\times\bbbr^d\to\bbbc$, and a sub-manifold
$\Omega\subseteq\bbbr^d$, the entries of the stiffness matrix
$G\in\bbbc^{\Idx\times\Jdx}$ are given by
\begin{align} \label{eqn:matrix_entry}
  g_{ij} &= \int_\Omega \varphi_i(x) \int_\Omega g(x,y) \psi_j(y) \,dy\,dx
\end{align}
for all $i\in\Idx$ and $j\in\Jdx$.
In this article, we focus on the kernel functions
\begin{align}\label{eqn:kernels}
  g(x,y) &= \frac{1}{4\pi \|x-y\|}, &
  g(x,y) &= \frac{\exp(\iota \kappa \|x-y\|)}{4\pi \|x-y\|}
\end{align}
corresponding to the Laplace and the Helmholtz equation.
Here $\iota\in\bbbc$ denotes the complex unit and $\kappa\in\bbbr$ is the
wave number.

Following the standard paradigm of finite element methods, the set
$\Omega$ is represented by a grid $\mathcal{G}$ consisting of triangles
$\Delta\in\mathcal{G}$.
Representing the entries of $G$ by
\begin{align*}
  g_{ij}
  &= \sum_{\Delta_\varphi\in\mathcal{G}}
     \sum_{\Delta_\psi\in\mathcal{G}}
            \int_{\Delta_\varphi} \varphi_i(x)
            \int_{\Delta_\psi} g(x,y) \psi_j(y) \,dy\,dx
\end{align*}
for all $i\in\Idx$ and $j\in\Jdx$, the task of computing these entries
reduces to the task of evaluating integrals on the few pairs
$(\Delta_\varphi,\Delta_\psi)\in\mathcal{G}\times\mathcal{G}$ of triangles
where both basis functions $\varphi_i$ and $\psi_j$ do not vanish.

\subsection{Sauter-Schwab quadrature}

Assume that $\Delta_\varphi,\Delta_\psi\in\mathcal{G}$ are given.
In order to evaluate the integral
\begin{equation*}
  \int_{\Delta_\varphi} \int_{\Delta_\psi} \varphi_i(x) g(x,y) \psi_j(y) \,dy\,dx,
\end{equation*}
we introduce the reference triangle
\begin{equation*}
  \widehat{\Delta} := \{ (s,t)\in\bbbr^2\ :\ 0\leq t\leq s\leq 1 \}
\end{equation*}
and affine maps $\Phi_\varphi:\widehat{\Delta}\to\Delta_\varphi$ and
$\Phi_\psi:\widehat{\Delta}\to\Delta_\psi$.
By the definition of surface integrals, we only have to evaluate
\begin{equation*}
  \sqrt{g_\varphi g_\psi}
  \int_{\widehat{\Delta}} \int_{\widehat{\Delta}}
    \varphi_i(\Phi_\varphi(\hat x)) g(\Phi_\varphi(\hat x),\Phi_\psi(\hat y))
    \psi_j(\Phi_\psi(\hat y)) \,d\hat y \,d\hat x,
\end{equation*}
where the Gramian determinants are denoted by
\begin{align*}
  g_\varphi &:= \det(D\Phi_\varphi^T D\Phi_\varphi), &
  g_\psi &:= \det(D\Phi_\psi^T D\Phi_\psi).
\end{align*}
For a polyhedral surface triangulation, both determinants are constant
and can be computed in advance.
Rewriting the integral with the transformed functions
\begin{align*}
  \hat\varphi_i &:= \varphi_i\circ\Phi_\varphi, &
  \hat\psi_j &:= \psi_j\circ\Phi_\psi, &
  \hat g(\hat x,\hat y) &:= g(\Phi_\varphi(\hat x),\Phi_\psi(\hat y))
\end{align*}
yields
\begin{equation}\label{eq:reference_integral}
  \int_{\widehat\Delta} \int_{\widehat\Delta}
    \hat\varphi_i(\hat x) \hat g(\hat x,\hat y) \hat\psi_j(\hat y)
    \,d\hat y\,d\hat x.
\end{equation}
For standard finite element discretizations, the transformed basis
functions $\hat\varphi_i$ and $\hat\psi_j$ are low-order polynomials
that can be integrated easily.
The transformed kernel function $\hat g$, on the other hand, poses
a challenge:
since typical kernel functions like the ones given in (\ref{eqn:kernels})
have singularities at $x=y$, we have to evaluate singular integrals.
The Sauter-Schwab quadrature rule \cite{SA92,SASC11} uses transformations
to remove these singularities, and the transformations depend on whether the
triangles $\Delta_\varphi$ and $\Delta_\psi$ are identical, share a common
edge, a common vertex, or are disjoint.
In the last case, the integrand is smooth and we can apply standard
quadrature.
In the first three cases, the integrand has a singularity on a
two-, one-, or zero-dimensional sub-manifold of the domain
$\widehat\Delta\times\widehat\Delta$ of integration.

If $\Delta_\varphi$ and $\Delta_\psi$ are disjoint, there is no singularity,
and we use the Duffy transformation
\begin{align*}
  T : [0,1]^2 &\to \widehat\Delta, &
      (s,t) &\mapsto (s,st),
\end{align*}
to write (\ref{eq:reference_integral}) in the form
\begin{equation*}
  \int_{[0,1]^4}
    \xi_1 \xi_3\,
    \hat\varphi_i(\xi_1,\xi_1\xi_2)\,
    \hat g(\xi_1,\xi_1\xi_2,\xi_3,\xi_3\xi_4)\,
    \hat\psi_j(\xi_3,\xi_3\xi_4) \,d\xi.
\end{equation*}
This integral can be approximated using standard tensor Gaussian
quadrature.

For the singular cases, \cite[\S 5.2]{SASC11} describes how the domain
$\widehat{\Delta}\times\widehat{\Delta}$ can be split and transformed
to obtain similar equations.
Since each of the cases requires us to split the domain into a different
number of subdomains, and since the transformations introduce additional
polynomial factors, the number of quadrature points depends on the case,
but not on anything else.

We can already see that Sauter-Schwab quadrature is particularly
attractive for GPGPUs:
the algorithm requires only the coordinates of the vertices of
$\Delta_\varphi$ and $\Delta_\psi$ and then performs a relatively large
number of operations, the evaluation of the kernel function and
the basis functions in all quadrature points, so we have
a fairly compute-intensive task.
It is also clear that for each of the four cases exactly the same
operations are carried out, only for different vertex data.
This means that the work required to handle large number of pairs of
triangles with the same type of singularity is ideally suited for
vectorization.

Our algorithm considers pairs of triangles, determines the
corresponding type of singularity, and creates work packages consisting
of triangle pairs with the same type.
Once the packages reach a given size, they are passed to the GPGPU
that performs the quadrature and returns the results to the CPU
that adds them to the appropriate coefficients in the matrix.

\subsection{Green cross approximation}

Computing all the entries of the matrix $G$ would lead to an
algorithm of quadratic complexity, and applying this algorithm to
complicated geometries with a large number of triangles would
involve a prohibitive computational cost.

This problem can be solved by replacing $G$ by a
\emph{data-sparse approximations}, e.g., constructed by the
fast multipole \cite{RO85,GRRO87,GRRO97} or the panel-clustering
\cite{HANO89} method.
Since we already have an algorithm for computing entries of $G$
at our disposal, we are looking for an approximation scheme that
makes use of these entries.
A well-known approach is the \emph{adaptive cross approximation}
(ACA) \cite{BERJ01} that uses a rank-revealing LR factorization to
construct blockwise low-rank approximations.
This procedure is adaptive, i.e., the number of iterations of its
main loop depends on the data that has to be approximated, so it
is not particularly well-suited for GPGPUs.

The recently developed \emph{Green-quadrature cross approximation}
(GCA) algorithm \cite{BOGO12,BOCH14}, on the other hand, can be
split into a computationally inexpensive adaptive part and a far
more expensive non-adaptive part.
We can afford to treat the adaptive part on the CPU, while GPGPUs
can take care of the bulk of the computation.

In order to derive the GCA method, we start with Green's representation
formula
\begin{equation*}
  u(x) = \int_{\partial\omega} g(x,z) \frac{\partial u}{\partial n}(z) \,dz
       - \int_{\partial\omega} \frac{\partial g}{\partial n(z)}(x,z) u(z) \,dy
\end{equation*}
for harmonic functions $u$ in a subdomain $\omega\subseteq\bbbr^3$.
If we pick $y\in\bbbr^3\setminus\bar\omega$, the restricted kernel function
$x\mapsto g(x,y)$ is itself harmonic, and applying Green's formula yields
\begin{equation*}
  g(x,y)
  = \int_{\partial\omega} g(x,z) \frac{\partial g}{\partial n(z)}(z,y) \,dz
  - \int_{\partial\omega} \frac{\partial g}{\partial n(z)}(x,z) g(z,y) \,dz.
\end{equation*}
If we make sure that $x$ and $y$ are sufficiently far from the boundary
$\partial\omega$ of the subdomain, both integrands appearing in this
equation are smooth and can be approximated by standard qua\-dra\-ture
rules, e.g., by a Gau\ss{} formula.
This allows us to replace the integrals by weighted sums
\begin{subequations}\label{eqn:green_apx}
\begin{align}
  g(x,y)
  &\approx \sum_{\nu=1}^m w_\nu g(x,z_\nu)
                  \frac{\partial g}{\partial n(z_\nu)}(z_\nu,y)\\
  &\quad - \sum_{\nu=1}^m w_\nu \frac{\partial g}{\partial n(z_\nu)}(x,z_\nu)
                  g(z_\nu,y),
\end{align}
\end{subequations}
where $(w_\nu)_{\nu=1}^m$ denotes the quadrature weights and $(z_\nu)_{\nu=1}^m$
denotes the quadrature points on the boundary $\partial\omega$.
Since the variables $x$ and $y$ are separated in this equation, it
can be used to construct low-rank approximations of the matrix $G$.

The GCA algorithm constructs an $\mathcal{H}^2$-matrix approximation
\cite{HAKHSA00,BOHA02,BO10} of $G$ in four steps:
\begin{enumerate}
  \item the index sets $\Idx$ and $\Jdx$ are split hierarchically
    into \emph{cluster trees},
  \item each cluster constructed in the first step is assigned
    a \emph{cluster basis},
  \item the index set $\Idx\times\Jdx$ is split hierarchically
    into a \emph{block tree}, and finally
  \item each block constructed in the third step is assigned either
    a \emph{near-field matrix} or a \emph{coupling matrix}.
\end{enumerate}
The first three steps are adaptive, but require only a moderate fraction
of the overall work.
The fourth step is computationally expensive, but non-adaptive, so this
step can be handled by GPGPUs.

A cluster tree $\ctI$ for an index set $\Idx$ is a tree of subsets of $\Idx$.
The root is the entire set $\Idx$ itself, and the sons $\sons(t)$ of a
non-leaf node $t\in\ctI$ correspond to a disjoint partition of their father.
The nodes of a cluster tree are called \emph{clusters}.
Cluster trees can be constructed very efficiently by geometrically
subdividing clusters \cite{GRHA02}.

A block tree $\ctIJ$ for the index set $\Idx\times\Jdx$ is a tree of
subsets $t\times s$ of $\Idx\times\Jdx$ with a \emph{row cluster}
$t\in\ctI$ and a \emph{column cluster} $s\in\ctJ$.
The sons $\sons(t\times s)$ of a non-leaf node are the Cartesian
products of the sons of $t$ and the sons of $s$.
The nodes of the block tree are called \emph{blocks}.
The block tree is used to identify submatrices that can be approximated
by low rank:
leaves are called \emph{admissible} if they correspond to submatrices
that can be approximated and \emph{inadmissible} otherwise.
Optimal block trees can be constructed using the cluster trees $\ctI$
and $\ctJ$ and an \emph{admissibility condition} that determines where
an approximation scheme can be applied \cite{GRHA02}.

In the case of GCA, we rely on a geometric admissibility condition:
clusters $t\in\ctI$ and $s\in\ctJ$ are associated with axis-parallel
\emph{bounding boxes} $B_t,B_s\subseteq\bbbr^3$ such that
\begin{align*}
  \supp \varphi_i &\subseteq B_t, &
  \supp \psi_j &\subseteq B_s &
  &\text{ for all } i\in t,\ j\in s.
\end{align*}
These bounding boxes can be constructed efficiently by merging
bounding boxes for the individual triangles of the grid $\mathcal{G}$.
Given an \emph{admissibility parameter} $\eta\in\bbbr_{>0}$, a pair
$(t,s)\in\ctI\times\ctJ$ of clusters is admissible, if
\begin{equation*}
  \max\{\diam(B_t), \diam(B_s)\} \leq \eta \dist(B_t,B_s)
\end{equation*}
holds.

Now we are in a position to apply (\ref{eqn:green_apx}).
Given a cluster $t\in\ctI$, we let $\delta_t := \diam(B_t)$ and construct
an axis-parallel box $\omega_t\subseteq\bbbr^3$ such that
$B_t\subseteq\omega_t$ and
\begin{equation*}
  \dist(B_t,\partial\omega_t) = \delta_t/2.
\end{equation*}
It is easy to see that this property implies
\begin{equation*}
  \dist(B_s,\partial\omega_t) \geq \delta_t/2
\end{equation*}
for all admissible pairs $(t,s)\in\ctI\times\ctJ$
(cf. \cite[Lemma~3]{BOCH14}), so the boundary $\partial\omega_t$
is sufficiently far from both $B_t$ and $B_s$ and we can apply
(\ref{eqn:green_apx}) to approximate the kernel function.

For $i\in t$ and $j\in s$ we obtain
\begin{align*}
  g_{ij} &\approx \sum_{\nu=1}^m
            \underbrace{\sqrt{w_\nu}
            \int_\Omega \varphi_i(x)
                g(x,z_\nu) \,dx}_{=:a_{t+,i\nu}}\\
  &\qquad\qquad \times \underbrace{\sqrt{w_\nu}
            \int_\Omega \psi_j(y)
                \frac{\partial g}{\partial n(z_\nu)}(z_\nu,y) \,dy}_{=:b_{ts+,j\nu}}\\
  &\quad -\sum_{\nu=1}^m
            \underbrace{\sqrt{w_\nu} \delta_t
            \int_\Omega \varphi_i(x)
                \frac{\partial g}{\partial n(z_\nu)}(x,z_\nu) \,dx}_{=:a_{t-,i\nu}}\\
  &\qquad\qquad \times \underbrace{\frac{\sqrt{w_\nu}}{\delta_t}
            \int_\Omega \psi_j(y)
                g(z_\nu,y) \,dy}_{=:b_{ts-,j\nu}}
\end{align*}
with the matrices $A_{t,+},A_{t,-}\in\bbbc^{t\times m}$ and
$B_{t,+},B_{t,-}\in\bbbc^{s\times m}$ including suitable scaling factors.
Introducing
\begin{align*}
  A_t &:= \begin{pmatrix} A_{t,+} & A_{t,-} \end{pmatrix}, &
  B_{ts} &:= \begin{pmatrix} B_{ts,+} & B_{ts,-} \end{pmatrix},
\end{align*}
we can write our result in the short form
\begin{equation}\label{eqn:green_matrix}
  G|_{t\times s} \approx A_t B_{ts}.
\end{equation}
It is important to note that the matrix $A_t$ depends only on the
row cluster $t$, but not on the column cluster $s$.
This is a very attractive property, since the number of clusters
is typically significantly smaller, frequently by a factor of 100
or more, than the number of blocks.

Experiments indicate that the rank $k=2m$ of the approximation
(\ref{eqn:green_matrix}) is frequently far greater than necessary.
In order to improve the efficiency, we apply the adaptive cross
approximation algorithm \cite{BERJ01} to the matrix $A_t$ and
obtain index sets $\tilde t\subseteq t$ and
$\tau\subseteq\{1,\ldots,k\}$ such that
\begin{equation}\label{eqn:aca}
  A_t \approx (A_t)_{t\times\tau} (A_t)_{\tilde t\times\tau}^{-1}
              (A_t)_{\tilde t\times k}.
\end{equation}
The right-hand side of this equation can be interpreted as an
``algebraic interpolation'' by introducing
\begin{equation*}
  V_t := (A_t)_{t\times\tau} (A_t)_{\tilde t\times\tau}^{-1}
\end{equation*}
and the ``restriction matrix'' $R_t\in\bbbr^{\tilde t\times t}$ satisfying
\begin{align*}
  R_t u &= u|_{\tilde t} &
  &\text{ for all } u\in\bbbc^t
\end{align*}
such that $R_t A_t = A_t|_{\tilde t\times k}$, and writing (\ref{eqn:aca})
in the short form
\begin{equation*}
  A_t \approx \mathfrak{I}_t A_t \qquad \text{ with } \qquad
  \mathfrak{I}_t := V_t R_t.
\end{equation*}
$\mathfrak{I}_t$ can be considered a Lagrange interpolation operator
mapping from $\bbbc^t$ into the range of $A_t$.
The columns of $V_t$ can be interpreted as Lagrange basis functions
satisfying
\begin{equation*}
  R_t V_t = (A_t)_{\tilde t\times\tau} (A_t)_{\tilde t\times\tau}^{-1} = I,
\end{equation*}
and the restriction matrix as the pointwise evaluation of the
interpolant.

Applying this interpolation operator to the submatrix $G|_{t\times s}$
and the approximation (\ref{eqn:green_matrix}) yields
\begin{equation*}
  \mathfrak{I}_t G|_{t\times s}
  \approx \mathfrak{I}_t A_t B_{ts}
  \approx A_t B_{ts}
  \approx G|_{t\times s},
\end{equation*}
i.e., the interpolation error is under control, and
\begin{equation*}
  \mathfrak{I}_t G|_{t\times s}
  = V_t R_t G|_{t\times s}
  = V_t G|_{\tilde t\times s},
\end{equation*}
i.e., only the rows for the indices $\tilde t\subseteq t$ are required
to construct the approximation.

\subsection{\texorpdfstring{$\mathcal{H}^2$-matrix}{H2-matrix}}

We can go one step further and apply the same arguments to the
column cluster $s$ instead of the row cluster $t$ to obtain an
index set $\tilde s\subseteq s$ and and interpolation operator
$\mathfrak{I}_s = V_s R_s$ such that
\begin{align*}
  G|_{t\times s} \mathfrak{I}_s^*
  &\approx G|_{t\times s}, &
  G|_{t\times s} \mathfrak{I}_s^*
  &= G|_{t\times\tilde s} V_s^*,
\end{align*}
and combining both interpolation operators yields
\begin{equation*}
  G|_{t\times s} \approx \mathfrak{I}_t G|_{t\times s} \mathfrak{I}_s^*
  = V_t G|_{\tilde t\times\tilde s} V_s^*.
\end{equation*}
The construction of the algebraic interpolation operators
$\mathfrak{I}_t$ and $\mathfrak{I}_s$ is adaptive, but has to be
performed only once per cluster.
When these operators are available, the approximation for a
block $t\times s$ can be constructed by evaluating the entries
of $G|_{\tilde t\times\tilde s}$, i.e., by performing quadrature.
We have already seen that GPGPUs are well-suited for this task,
and only a very slight modification of our implementation is
necessary.

We can improve the efficiency of the GCA scheme even further by
constructing \emph{nested} cluster bases:
if $t\in\ctI$ is a leaf, we proceed as before.
Otherwise we first construct the interpolation operators for its
sons.
For the sake of simplicity, assume that there are two sons $t_1,t_2$
and that corresponding subsets $\tilde t_1\subseteq t_1$,
$\tilde t_2\subseteq t_2$ and matrices $V_{t_1},V_{t_2}$ have already
been constructed.
If $(t,s)\in\ctI\times\ctJ$ is admissible, the same holds for
$(t_1,s)$ and $(t_2,s)$, and we have
\begin{equation*}
  G|_{t\times s}
  = \begin{pmatrix}
      G|_{t_1\times s}\\
      G|_{t_2\times s}
    \end{pmatrix}
  \approx \begin{pmatrix}
      V_{t_1} G|_{\tilde t_1\times s}\\
      V_{t_2} G|_{\tilde t_2\times s}
    \end{pmatrix}
  = \begin{pmatrix}
      V_{t_1} & \\
      & V_{t_2}
    \end{pmatrix} G|_{\hat t\times s}
\end{equation*}
with $\hat t := \tilde t_1 \cup \tilde t_2$.
Applying cross approximation to the reduced matrix
$\widehat{A}_t := A_t|_{\hat t\times k}$ instead of $A_t$ yields a
new interpolation operator with an index set $\tilde t\subseteq\hat t$
and a matrix $\widehat{V}_t$ such that
\begin{equation*}
  G|_{t\times s}
  \approx \begin{pmatrix}
            V_{t_1} & \\
            & V_{t_2}
          \end{pmatrix} G|_{\hat t\times s}
  \approx \begin{pmatrix}
            V_{t_1} & \\
            & V_{t_2}
          \end{pmatrix} \widehat{V}_t G|_{\tilde t\times s},
\end{equation*}
and we can define
\begin{equation*}
  V_t := \begin{pmatrix}
           V_{t_1} & \\
           & V_{t_2}
         \end{pmatrix} \widehat{V}_t
\end{equation*}
to obtain the by now familiar form
\begin{equation*}
  G|_{t\times s} \approx V_t G|_{\tilde t\times s}.
\end{equation*}
Since $\hat t$ is typically a significantly smaller set than
$t$, this recursive procedure can be significantly faster than
the direct approach.
The storage requirements are also reduced, since we have to
store $V_t$ only for leaf clusters and can use the smaller matrices
$\widehat{V}_t$ for all other clusters.
The resulting structure is known as an \emph{$\mathcal{H}^2$-matrix}
\cite{HAKHSA00,BOHA02,BO10}, the algebraic counterpart of fast
multipole methods \cite{RO85,GRRO87}, and it is possible to prove
that the storage requirements and computational complexity of
$\mathcal{H}^2$-matrices only grow linearly with the number of
degrees of freedom and the rank.
This makes them very attractive for large-scale computations.

\section{Implementation}

Although the previously mentioned computations are well-suited for
GPGPUs in principle, we have to overcome a number of challenges
in order to make the most of the available recources.

To set the stage, we consider a rough sketch of how a typical
GPGPU works.
It usually consists of $n_{mp}$ \emph{multiprocessors}, each equipped
with $n_{pe}$ arithmetic-logical \emph{processing elements}.
All processing elements of a given multiprocessor can either perform
the same operation (addition, multiplication, comparison, jump, ...)
or no operation in a given clock cycle.
The SIMT (single instruction, multiple threads) programming model
splits a computation into a number of \emph{threads}, sequences of
instructions that can be executed by the processing elements.
Each thread is associated with an instruction counter that decides which
instruction it has to execute next.

Since all processing elements of a multiprocessor have to perform the
same operation in each cycle, the threads are bound together to form
\emph{warps}, each warp consisting of $n_{pe}$ threads, and these warps
are assigned to one of the multiprocessors.

In a clock cycle, each of the multiprocessors chooses a warp for
execution, chooses an instruction, and performs the corresponding
operation for all threads in the warp whose instruction counter
matches the chosen instruction.

In order to take full advantage of all processing elements, it is
therefore crucial that all threads, or at least all threads in each
warp, execute the same instruction most of the time.
Usually GPGPUs are ``smart enough'' to recover from threads that
diverge only for a few cycles and then merge again, but it is our
responsibility as programmers to keep these instances of instruction
divergence as brief as possible.

Regarding our algorithm, the two most important problems are that
\begin{itemize}
  \item near-field matrix blocks $G|_{t \times s}$ usually consist of
        a mix of all four cases of the Sauter-Schwab quadrature, each
        requiring a different number of quadrature points, and that
  \item a single block $G|_{t \times s}$ or $G|_{\tilde t \times \tilde s}$
        is too small to take advantage of the thousands of processing
        elements provided by a modern GPGPU on its own.
\end{itemize}
The first issue is problematic since it means that mixing threads
handling different cases within the same warp would lead to instruction
divergence, the time for the execution of the warp would be (at best)
determined by the longest-running thread.

The second issue poses a challenge since initiating the execution
of a number of threads on a GPGPU typically requires a significant
amount of bookkeeping on one hand and communication via a slow bus
system on the other one.
Some older GPGPUs even can only execute the same code on all
multiprocessors, so running 10 threads on a GPGPU with 1\,024
processing elements will lead to 1\,014 processing elements
doing nothing.

To address the first issue, we keep track of all integrals that
appear in the course of the matrix assembly.
Depending on the quadrature case, the indices of the triangles
and the memory cell the integral belongs to in the matrix are
be saved in individual lists for all four cases.
If we launch threads on the GPGPU for all integrals in one of these
lists, no instruction divergence can occur anymore, since all threads
in one of the lists execute exactly the same instructions in exactly
the same order.

A welcome side-effect of this approach is that we can control the size
of these lists.
This also addresses the second issue mentioned before:
there is no longer a need for handing only a small number of integrals
to the GPGPUS, we can choose the maximal size of the lists to strike
a good balance between communication and bookkeeping overhead and
ease of implementation.
Experiments show that the size should be several megabytes to obtain
a reasonable performance.

This algorithmic realization has some drawbacks as well:
we have to allocate extra memory for the lists, but since the user can
choose the size of the lists and since the storage can be re-used after
one computation of a list has been finished, this is not a major drawback.

\begin{figure}
\label{img:single_multithreaded}
\begin{center}
 \subfigure[GPGPU occupancy with a single worker thread]{
   \includegraphics[width=0.45\textwidth]{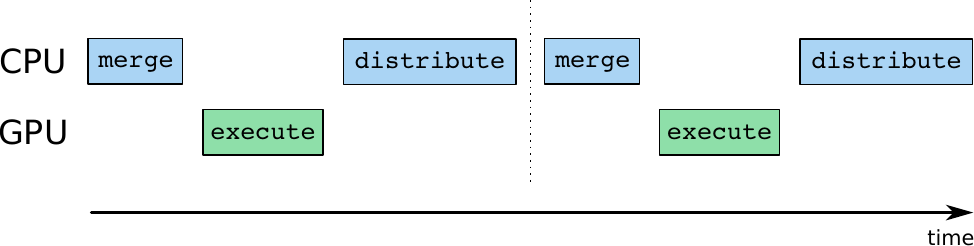}
  }
 
 \subfigure[GPGPU occupancy with two worker threads]{
   \includegraphics[width=0.45\textwidth]{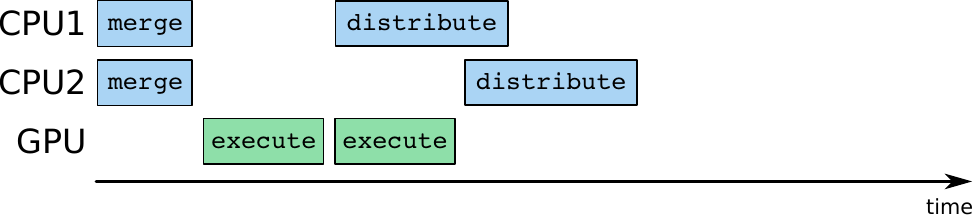}
  }
\end{center}
\caption{Schematic difference of a single worker threads vs.
  two worker threads supplying a GPGPU with work items.}
\end{figure}

On the other hand, after the computations on the GPGPU have finished,
the integrals have to be added to their final position in the stiffness
matrix $G$.
These memory operations have to be performed by the CPU, since we
assume that only the CPU has a sufficient amount of memory at its
disposal to store the matrix, and in a single-threaded program the GPGPU
cannot do any further computations because the CPU is not able to provide
work to the GPGPU while copying data.
Consequently, we propose to use multiple CPU threads to overcome this
problem (cf. figure \ref{img:single_multithreaded}).

Of course, it would also be beneficial to let the CPU cores compute
integrals, since they also contain SIMD units.
Here we focus on an approach that has a single thread, in the following
called the \emph{master thread}, traverse the block tree and fill the
lists with work items, while other CPU threads, called \emph{worker threads}
have the task to prepare the data in the lists for the GPGPU computation,
launch appropriate threads on the GPGPU, and distribute the results back
to the stiffness matrix.
Essentially, we use a simple task scheduler following the producer-consumer
pattern.

The implementation is based on the hierarchical matrix library
\emph{H2Lib}\footnote{Available at \url{http://www.h2lib.org}}
written in the C programming language.
For the realization of the task scheduler on the main processor, we
employ OpenMP to create the master thread and a number of wor\-ker threads.
The computations of the GPGPU are carried out using the OpenCL standard,
which allows us to execute the quadrature code on any platform that
implements OpenCL, not just on GPGPUs.
Of course, our approach can also be applied with different threading
libraries and different GPGPU APIs like CUDA.

In order to keep the amount of data that has to be communicated to
the GPGPU small, we use a trick:
in a first step, all integrals in a block $G|_{t\times s}$ are treated
as if they were regular, i.e., as if they corresponded to disjoint
triangles.
This allows us to characterize the entire block just by the sets $t$
and $s$, and we avoid having to send all indices of the Cartesian
product $t\times s$ to the GPGPU.
Of course, this will lead to wrong results if some of the integrals
are singular, and so we recompute these problematic integrals in a
second phase and fix the corresponding matrix entries.
Although this means that some integrals will be computed ``twice'',
the savings afforded by the reduction in communication more than
make up for this overhead.

We split the assembly of the stiffness matrix into the following parts:
\begin{enumerate}
 \item Set up all algebraic interpolation operators $\mathfrak{I}_t$
       and $\mathfrak{I}_s$ for all subsets $t \subseteq \Idx$ and
       $s \subseteq \Jdx$ on the CPU, since this cannot be properly
       vectorized.
 \item \label{collectlist}
       Split the program execution into the execution of a
       master thread and several worker threads.
       The master thread traverses the block tree and adds each block
       to the list of regular integrals to be computed.
       
       As soon as the total size of one of these lists exceeds a given limit,
       the list is marked as ``ready for execution'' and handed over to the
       worker threads.
       This is depicted in algorithm \ref{alg:assembly}.
 \item Once a list has been marked as ``ready for execution'', a worker thread
       can fetch this list (cf. algorithm \ref{alg:scheduler})
       and process this data package (see algorithm \ref{alg:execute_gpu}).
 \item Once a work item has been processed on the GPGPU, the results
       need to be copied to their proper location in the stiffness
       matrix by the CPU.
       In case a block contains non-disjoint pairs of triangles, they
       are now added to the list for their respective case and handled
       as in (\ref{collectlist}) which is also displayed in algorithm
       \ref{alg:distribute_disjoint}.
\end{enumerate}

\begin{algorithm}
\begin{algorithmic}
\Procedure{Add\_block}{b}
  \If {\Call{sizeof}{$\ell$} + \Call{sizeof}{b} $<$ maxsize}
    \State \Call{Add\_to\_list}{$\ell$,b}
  \Else
    \State \Call{Enqueue\_list}{$\ell$}\;
    \State $\ell$ = \Call{New\_list}{maxsize}\;
    \If {\Call{sizeof}{$\ell$} + \Call{sizeof}{b} $<$ maxsize}
      \State \Call{Add\_to\_list}{$\ell$,b}
    \Else
      \State (b1,b2) $\gets$ \Call{Split\_block}{b}
      \State \Call{Add\_block}{b1}
      \State \Call{Add\_block}{b2}
    \EndIf
  \EndIf
\EndProcedure

\State

\Procedure{Assemble\_GPU}{$G$}
\ForAll {matrix blocks $b \in \left\{G_{|t \times s}, 
   G_{|\hat t \times \hat s}\right\}$}
  \State \Call{Add\_block}{b}
\EndFor
\EndProcedure
\end{algorithmic}
\caption{Assembly of a boundary integral operator $G$. 
   The parameters of all near-field and far-field blocks 
   are initially added to the list of disjoint triangles.
   The procedure {\sc Add\_block} takes care that these lists
   do not become too big and are queued for execution by the
   scheduling algorithm once they reach a given size.}
\label{alg:assembly}
\end{algorithm}

\begin{algorithm}
\begin{algorithmic}
 \Procedure{Schedule\_tasks}{}
   \While{$\ell \gets$ \Call{Dequeue\_list}{}}
     \If {$\ell$ $\neq$ empty}
       \If {\Call{Affinity}{$\ell$} $=$ GPU}
	 \State \Call{Execute\_GPU}{$\ell$}
       \Else
	 \State \Call{Execute\_CPU}{$\ell$}
       \EndIf
     \Else
       \State \Call{Sleep}{}
     \EndIf
   \EndWhile
 \EndProcedure
\end{algorithmic}
\caption{Basic sketch of the scheduler algorithm that is used.
   As long as new lists are available, each thread fetches
   a list and either executes it on the CPU or on the GPGPU
   depending on its affinity.
   When running out of lists to be processed, a thread goes to sleep
   for a short period of time or terminates if no more lists are to be
   expected.}
\label{alg:scheduler}
\end{algorithm}

\begin{algorithm}
\begin{algorithmic}
 \Procedure{Distribute\_results\_disjoint}{$\ell$, data}
  \ForAll {$G_{|t \times s} \in \ell$}
    \State \Call{Copy\_results}{data, $G_{|t \times s}$}

    \If{$G_{|t \times s}$ contains non-disjoint pairs of triangles}
      \ForAll {$(i,j) \in G_{|t \times s}$}
        \State $v \gets$ \Call{Number\_of\_common\_vertices}{t, s}
        \If{$v > 0$}
          \State \Call{Add\_entry\_to\_list}{$\ell_v$, $(i,j)$}
        \EndIf
      \EndFor
    \EndIf
  \EndFor
 \EndProcedure
\end{algorithmic}
\label{alg:distribute_disjoint}
\caption{Accumulation for regular integrals.
   If a block also contains non-disjoint pairs of triangles,
   corresponding to singular integrals, they are now added to the list
   of their respective case and processed in a similar way.}
\end{algorithm}

\begin{algorithm}
\begin{algorithmic}
 \Procedure{Execute\_GPU}{$\ell$}
   \State data $\gets$ \Call{Merge\_data}{$\ell$}
   \State \Call{Invoke\_GPU\_kernel}{data}
   \State \Call{Distribute\_results}{$\ell$, data}
   \State \Call{Cleanup}{data}
 \EndProcedure
\end{algorithmic}
\label{alg:execute_gpu}
\caption{Executing a list $\ell$ on the GPGPU.
   The parameters of the individual integrals have to be copied into
   contiguous chunks of memory to be transferred to the GPGPU before
   the threads can be launched on the GPGPU.
   After completion, the results have to be copied back to main memory
   and distributed in the matrix $G$.}
\end{algorithm}

\section{Numerical experiments}

In this part we want to demonstrate the performance of our GPGPU 
algorithm for setting up boundary integral operators arising in
the boundary element method.

For the following experiments we consider the solution of an interior 
Dirichlet problem for the Laplace equation in 3D.
Let $f$ be a harmonic function in a bounded domain 
$\Omega \subset \mathbb{R}^3$ and assume that the Dirichlet values
$f|_{\partial\Omega}$ are given.
In the context of the boundary element method, this translates to
the boundary integral equation
\begin{equation}
\label{eqn:varform_laplace}
 \int_{\del \Omega} g(x,y) \frac{\del f}{\del n}(y) \,dy
   = \frac{1}{2} f(x) + 
     \int_{\del \Omega} \frac{\del g}{\del n(y)}(x,y) f(y) \,dy
\end{equation}
for (almost) all $x\in\partial\Omega$.
We are looking for the Neumann values $\frac{\partial f}{\partial n}$.
Here the kernel function
\begin{equation*}
  g(x,y) := \frac{1}{4\pi\|x-y\|}
\end{equation*}
is the fundamental solution of the Laplace operator in 3D.

A Galerkin discretization of this equation with piece-wise constant
basis functions $( \varphi_i )_{i \in \Idx}, ( \psi_j )_{j \in \Jdx}$ leads
to a linear system
\begin{equation*}
  V \alpha = \left( \frac{1}{2}I + K \right) \beta,
\end{equation*}
where the coefficients of the matrices $V$ and $K$ are given by
equation \eqref{eqn:matrix_entry} with the kernel
functions $g$ and $\frac{\partial g}{\partial n(y)}$, respectively.
The values of $f$ and $\frac{\partial f}{\partial n}$ are represented in
the bases as $f = \sum_{j \in \Jdx} \beta_j \psi_j$ and
$\frac{\partial f}{\partial n} = \sum_{i \in \Idx} \alpha_i \varphi_i$,
respectively. Since the matrix $V$ is symmetric and positive
definite, we can employ a conjugate gradient method \cite{HEST52} to
solve the linear system.

The most time-consuming part of the whole computation is the setup
of the matrices $V$ and $K$, which is carried out mostly by the GPGPU.
Due to the symmetry of $V$, we only have to compute its lower
triangular part.
The setup times for $K$ are roughly twice those for the matrix $V$,
not only due to the fact that the entire matrix $K$ is constructed,
but also since the kernel function of $K$ requires more operations
than the one for $V$.

In order to verify the correctness of our proposed method, we solve
the system with the following three harmonic functions:
\begin{gather*}
  f_1(x) = x_1^2 - x_3^2,\qquad
  f_2(x) = g(x, (1.2, 1.2, 1.2)),\\
  f_3(x) = g(x, (1.0, 0.25, 1.0)).
\end{gather*}
As a simple test domain we use the unit sphere
\begin{equation*}
  \Omega_s=\{x\in\bbbr^3\ :\ \|x\|_2=1\},
\end{equation*}
approximated by a triangulation constructed by regularly subdividing the
eight triangles forming the double pyramid
$\Omega_p=\{x\in\bbbr^3\ :\ \|x\|_1=1\}$ into subtriangles and
projecting the vertices to $\Omega_s$.

\paragraph*{CPU algorithm}

\begin{figure}
\begin{center}
   \includegraphics[width=0.45\textwidth]{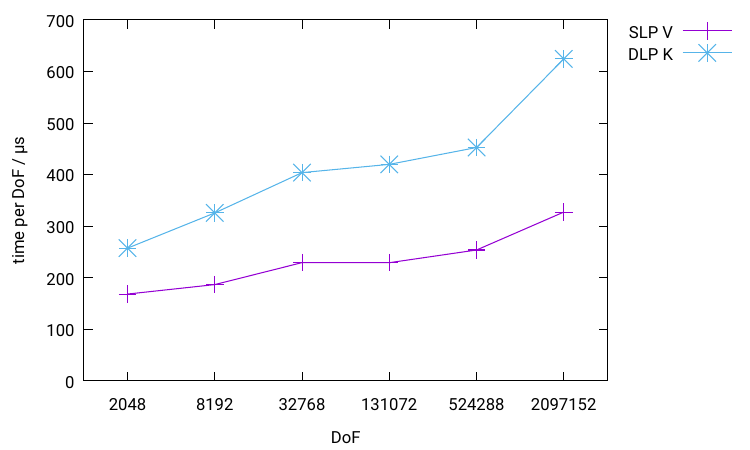}
\end{center}
\caption{Setup time of the boundary integral operators on different
         resolutions of the unit sphere.
         Only CPU threads are employed to calculate the integrals.}
\label{img:cpu}
\end{figure}

To have a reference point for our further experiments, we first
present some results computed entirely on the CPU.
Instead of traversing the block tree and keeping account of the
the arising matrix sub-blocks to be computed, the CPU algorithm
directly computes these matrix entries.
The algorithm was executed on a Intel Core i7-3820 processor
with 4 cores and hyper-threading enabled, which allows us to use up
to 8 threads on the CPU.
Consequently, we want to use all 8 threads to execute the algorithm
on the CPU in order to keep pace with the GPGPUs later on.
The quadrature routines employ vectorization based on AVX.
The results can be seen in Figure~\ref{img:cpu}.
The timings do not include the setup time for the algebraic interpolation 
operators, since these are always precomputed by a CPU algorithm.
In this and all later experiments we scale the runtime by the number
of degrees of freedom in order to compare the values across a wide range of
resolutions of the geometry.
We can see in Figure~\ref{img:cpu} that the runtime on the main processor is
between 150 and 650 microseconds ($\mu$s) for both the SLP operator $V$ and
the DLP operator $K$ for all resolutions.

The increase in time per degree of freedom is due to higher accuracies
for the GCA algorithm:
as we refine the grid, the condition number of $V$ grows, and we have
to approximate the matrix more accurately in order to preserve the
convergence rate of the original (uncompressed) boundary element method.
The parameters of the GCA approximation scheme are chosen as in
\cite{BOCH14} to ensure $\mathcal O(h)$ convergence of the
Neumann values and to keep them close to the discretization error. 
The increase in accuracy leads to the poly-logarithmic growth  observed
in Figure~\ref{img:cpu}.

\paragraph*{List size}

\begin{figure*}
\begin{center}
 \subfigure[Runtime for $V$ and $K$ on GeForce GTX 580]{
   \label{img:listsize:a}
   \includegraphics[width=0.47\textwidth]{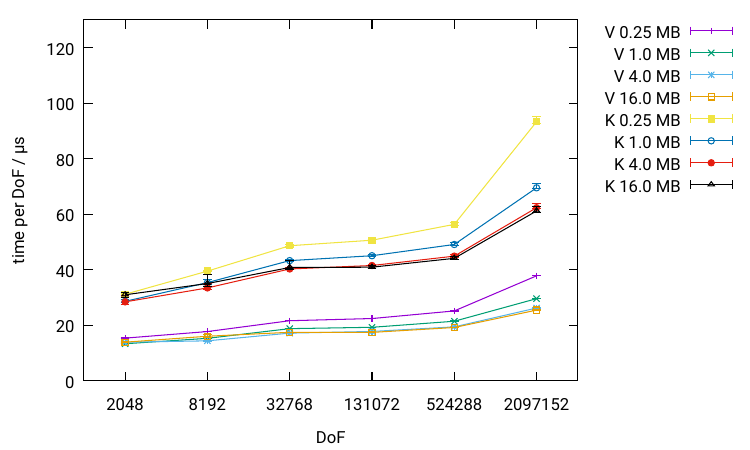}
   }
 \subfigure[Runtime for $V$ and $K$ on GeForce GTX 680]{
   \label{img:listsize:b}
   \includegraphics[width=0.47\textwidth]{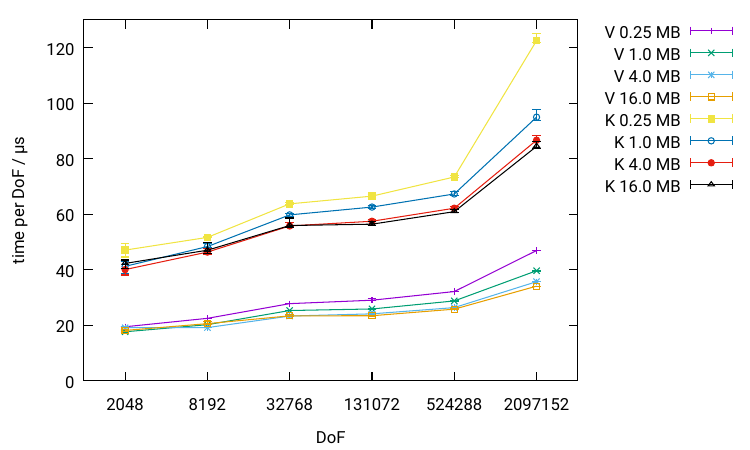}
   }
\end{center}
\caption{Setup time of the boundary integral operators for varying 
         list sizes on different resolutions of the unit sphere.
         An NVIDIA GeForce GTX 580 was used to compute the SLP operator $V$
         and the DLP operator $K$ in \ref{img:listsize:a}.
         Likewise an NVIDIA GeForce GTX 680 was used to compute the SLP
         and DLP operators in \ref{img:listsize:b}.}
\label{img:listsize}
\end{figure*}

Now we consider the performance of our GPGPU algorithm, starting
with the influence of the size of lists that are collected before
being sent to the GPGPUs.
It seems obvious that a smaller size automatically corresponds to
more interactions between CPU and GPGPUS, and since each transfer
between CPU and GPGPU involves some overhead, it is not surprising
that smaller list sizes lead to longer execution times of our algorithm.
The setup times for different list sizes and different resolutions
are given in Figure~\ref{img:listsize}.

Due to the non-deterministic behaviour of the schedulers both on the CPU
and the GPGPU, we have run the computations five times on the GPGPU and
noted the minimal, maximal, and average runtimes of the algorithm. 
These are depicted by the error-bars in Figure~\ref{img:listsize} and
the following ones.
We can see that the deviation is larger for very small list sizes
and negligible for larger ones.
The main conclusion drawn from figure~\ref{img:listsize} is that the
list size should be chosen somewhere around 4--16 MB to minimize the
runtime. However this is not the best choice in all cases.
In general there is no best choice for all degrees of freedom 
for this parameter, but 8 MB seems to be a good empirical value.

%
%
\begin{remark}
  At the first glance, the results from figure \ref{img:listsize} appear
  surprising because the GeForce GTX 680 card achieves a lower 
  performance than its architectural predecessor, the GeForce GTX 580.
  But in fact, this can be explained by changes in the GPGPU's 
  internal structure, especially for double precision computations.
  The theoretical peak performance of the GeForce GTX
  680 is given by 128.8 GFlops and the peak performance of the GeForce
  GTX 580 is given as 197.6 GFlops in double precision.
  Therefore we can expect the GeForce GTX 580 to be faster than the
  GeForce GTX 680 by a factor of 1.5 which is confirmed by our results.

  In comparison, the current GeForce GTX 1080 card is reported to
  reach 277 GFlops in double precision computations, so we expect it
  to be only about 50\% faster than the GTX 580.
\end{remark}

\paragraph*{Worker threads}

Since there is always a portion of time in between two successive
kernel calls on the GPGPU where the CPU has to perform some
data management operations, the utilization of the GPGPU will not be
near its maximum.
Therefore our implementation of the algorithm has a parameter
controlling the number of CPU worker threads spawned per GPGPU to
supply the GPGPU with work.
We now investigate the effect of the number of worker threads
on the overall performance of our approach.

\begin{figure*}
\begin{center}
 \subfigure[Runtime for $V$ and $K$ on GeForce GTX 580]{
   \label{img:workerthreads:a}
   \includegraphics[width=0.47\textwidth]{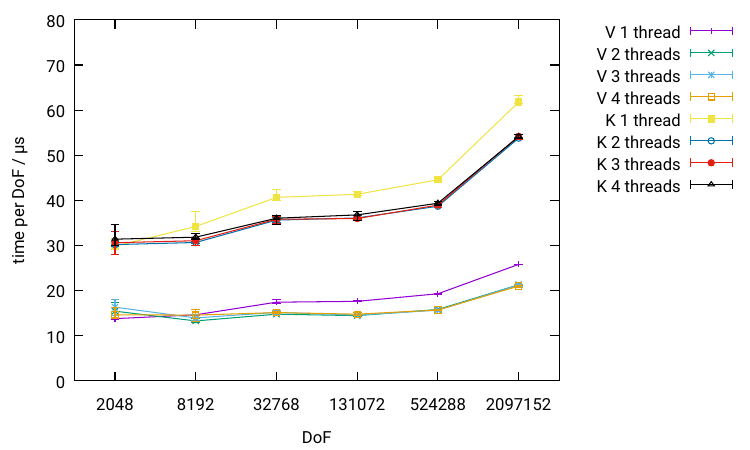}
   }
 \subfigure[Runtime for $V$ and $K$ on GeForce GTX 680]{
   \label{img:workerthreads:b}
   \includegraphics[width=0.47\textwidth]{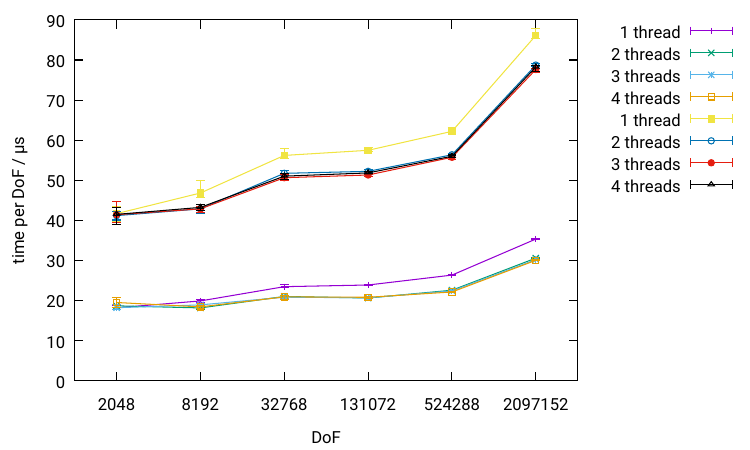}
   }
\end{center}
\caption{Setup time of the boundary integral operators for varying 
         number of worker threads per GPGPU on different resolutions 
         of the unit sphere. The list size was chosen constant as 
         8 MB.
         An NVIDIA GeForce GTX 580 was used to compute the SLP matrix $V$
         and the DLP matrix $K$ in \ref{img:workerthreads:a}.
         Likewise an NVIDIA GeForce GTX 680 was used to compute the
         SLP and DLP matrices in \ref{img:workerthreads:b}.}
\label{img:workerthreads}
\end{figure*}

In Figure~\ref{img:workerthreads} we can see a significant reduction
of the run\-time when two or more worker threads are employed per GPGPU.
It is not necessary to use more than two threads per GPGPU.
Indeed using more than two threads might have a negative impact
on the performance if not enough hardware threads are available
to supply the GPGPUs with sufficient work or the memory bandwidth
of the CPU might be exhausted at some point.
We can also observe that the deviation in runtime increases for
smaller resolutions of the geometry and decreases for bigger resolutions
as more worker threads are employed. 
This might be due to the non-deterministic behaviour of CPU thread
scheduling by the operating system.
Additionally, for very small numbers of unknowns using more worker
threads can also influence the runtime in a negative way because
there is not enough work to distribute and some threads might be idle.

\paragraph*{Multi-GPU setup}

\begin{figure*}
\begin{center}
 \subfigure[Runtime for $V$ on GeForce GTX 580 and 680]{
   \includegraphics[width=0.47\textwidth]{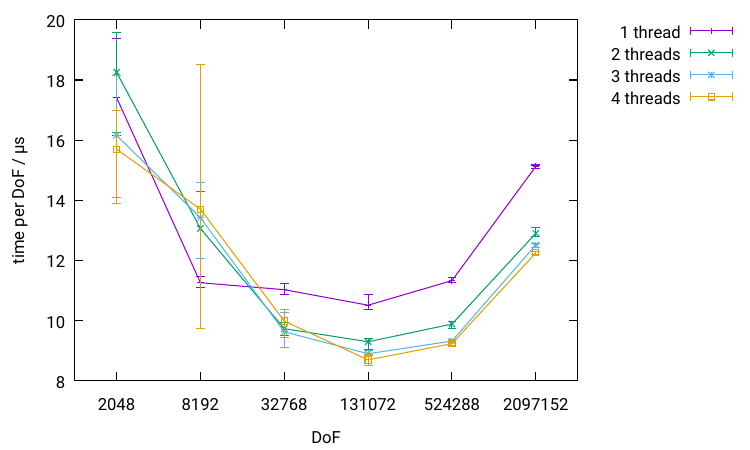}
   }
 \subfigure[Runtime for $K$ on GeForce GTX 580 and 680]{
   \includegraphics[width=0.47\textwidth]{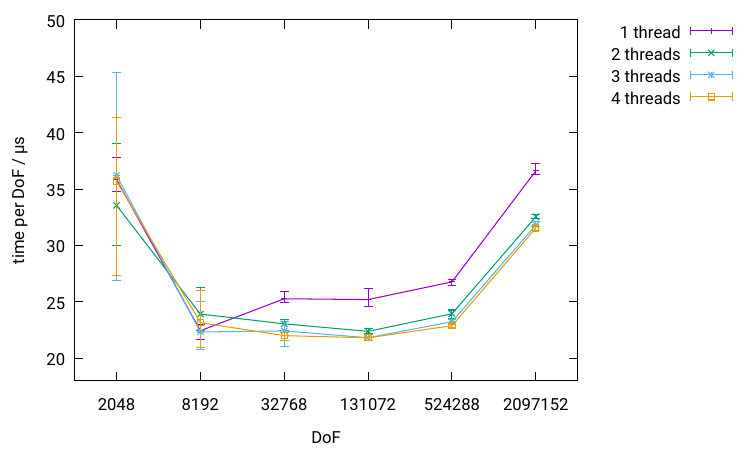}
   }
\end{center}
\caption{Setup time of the boundary integral operators using 
         both GeForce GTX 580 and GeForce GTX 680 
         to compute the integrals.
         The list size is kept constant at 8 MB and two worker
         threads are being employed per GPGPU.}
\label{img:multigpu}
\end{figure*}

In the design of our algorithm, we took into account that a
system can have more than just one GPGPU available.
A single worker thread is always directly connected with a
single GPGPU and cannot supply work to a different GPGPU.
With this concept it is very straightforward to employ all GPGPUs in
a system to work on the same problem.
On the other hand, all GPGPUs are connected by the same PCIe bus system
in typical computer systems, and it is not quite clear if the 
total amount of data that is being transferred over this bus can 
be handled without any delay.
We investigate this behaviour in the next experiment, where we employ
both a GeForce GTX 580 and GeForce GTX 680 at the same time.
In this context we are interested in how the runtime of our algorithm
scales when using both cards simultaneously.
Figure~\ref{img:multigpu} illustrates the effect of the multi-GPU
setup.

At least for this two-GPU setup, our approach appears to scale well.
According to our previous remark about the peak performances of both
GPGPUs we can only expect the runtime for the two-GPU setup of our algorithm
to be smaller by a factor of 1.7 compared to the runtime of the
GeForce GTX 580 or smaller by a factor of 2.5 compared to the
runtime of the GeForce GTX 680.
Indeed our experiments show similar factors, and we can say that our
scheduler handles the two-GPU setup quite well.

The scaling behaviour of our algorithm on more than two GPGPUs is an 
interesting topic and will be part of future research.

\begin{table*}
\centering
 \caption{Runtimes for CPU and GPGPU algorithms to setup SLP operator $V$.
      List size is kept constant at 8 MB and 2 worker threads are
      employed.} {
  \begin{tabular}{r|r|r|r|r|r}
  DoFs & $\mathfrak{I}_t, \mathfrak{I}_s$ & CPU & GTX 580 & 
  GTX 680 & GTX 580 + 680\\
  \hline
  2048 & $<$0.1 s & 0.3 s & $<$0.1 s & $<$0.1 s & $<$0.1 s\\
  8192 & 0.1 s & 1.5 s & 0.1 s & 0.2 s & 0.1 s\\
  32768 & 0.4 s & 7.5 s & 0.5 s & 0.7 s & 0.3 s\\
  131072 & 1.5 s & 30.0 s & 1.9 s & 2.7 s & 1.2 s\\
  524288 & 6.2 s & 133.0 s & 8.3 s & 11.8 s & 5.2 s\\
  2097152 & 62.0 s & 686.2 s & 44.6 s & 64.1 s & 27.1 s
  \end{tabular}}
\label{tab:final_laplace_slp}
\end{table*}

\begin{table*}
\centering
 \caption{Runtimes for CPU and GPGPU algorithms to setup DLP operator $K$.
      List size is kept constant at 8 MB and 2 worker threads are
      employed.} {
  \begin{tabular}{r|r|r|r|r|r}
  DoFs & $\mathfrak{I}_t, \mathfrak{I}_s$ & CPU & GTX 580 & 
  GTX 680 & GTX 580 + 680\\
  \hline
  2048 & $<$0.1 s & 0.5 s & $<$0.1 s & $<$0.1 s & $<$0.1 s\\
  8192 & 0.1 s & 2.7 s & 0.3 s & 0.4 s & 0.2 s\\
  32768 & 0.5 s & 13.2 s & 1.2 s & 1.7 s & 0.8 s\\
  131072 & 1.7 s & 55.0 s & 4.7 s & 6.8 s & 2.9 s\\
  524288 & 6.7 s & 237.3 s & 20.3 s & 29.5 s & 12.5 s\\
  2097152 & 70.2 s & 1309.0 s & 112.8 s & 165.1 s & 68.2 s
  \end{tabular}}
\label{tab:final_laplace_dlp}
\end{table*}

Tables~\ref{tab:final_laplace_slp} and \ref{tab:final_laplace_dlp}
list the absolute computing times for our algorithm and also include
the setup time for the algebraic interpolation operators for the sake of
completeness, although they are computed entirely on the CPU.
We observe that the time required to set up the interpolation operators
is roughly the same as the time required by the setup of the near-field
and coupling matrices on both GPGPUs.

We could speed up computation of the interpolation operators as
well, either by using SIMD instructions and multiple cores on the CPU or
with the help of GPGPUs.

More important is the time needed to solve the three linear systems.
We used a simple conjugate gradient method which needed 260 steps on
the finest mesh and took about 1400 seconds for each right-hand side.
At this point a good preconditioner would be very desirable.
The \emph{H2Lib} package offers $\mathcal{H}^2$-Cho\-les\-ky and
$\mathcal{H}^2$-LU decompositions, but unfortunately these are not
yet parallelized.
Hence a comparison with a sequential algorithm for constructing
a preconditioner would not be fair.

\paragraph*{Higher quadrature orders}

All experiments so far have been conducted with a basic one dimensional
Gaussian quadrature formula with 3 points for Duffy
transformations and with 5 points for the Sauter-Schwab quadrature formulas.
In some applications this quadrature order might not be sufficient if
the geometries become less smooth.
Therefore we also present some results for a basic Gaussian quadrature
formula with 4(6) quadrature points in Table~\ref{tab:q4}.
Since the unit sphere is a very smooth geometry, such a high quadrature
order is not really necessary, but we can see that the GPGPUs benefit
from the higher order, since their efficiency increases with the amount
of work required for a single integral.
Hence the speedup factor compared to the CPU implementation rises from 25.3
to 29.8 for the single layer potential and from 19.2 to 22.3 for the double
layer potential.

\begin{table*}
\centering
 \caption{Runtimes for CPU and GPGPU algorithms to setup SLP $V$ and DLP
      operator $K$ for basic quadrature rules using 4(6) points.
      List size is kept constant at 8 MB and 2 worker threads are
      employed.
      For the last line, the GCA parameter $\delta=2$ was chosen
      instead of $\delta=1$ to make the problem fit into memory.} {
  \begin{tabular}{r|r|r|r|r}
  DoFs & CPU SLP & GTX 580 + 680 SLP & CPU DLP & GTX 580 + 680 DLP\\
  \hline
  2048 & 1.0 s & $<$0.1 s & 1.5 s & 0.1 s\\
  8192 & 4.8 s & 0.2 s & 7.7 s & 0.4 s\\
  32768 & 21.6 s & 0.7 s & 39.0 s & 1.8 s\\
  131072 & 86.8 s & 2.8 s & 162.1 s & 7.3 s\\
  524288 & 368.5 s & 12.0 s & 702.0 s & 31.3 s\\
  2097152 & 1439.3 s & 48.3 s & 2936.2 s & 131.9 s
  \end{tabular}}
\label{tab:q4}
\end{table*}

\paragraph*{Helmholtz equation}

As a final example we consider a slightly different problem
setting: the Helmholtz equation in its BEM formulation.
In this case we would like to solve the exterior Dirichlet problem,
given by
\begin{align*}
 \Delta u(x) + \kappa^2 u(x) &= 0, & x \in \mathbb{R}^3 
    \setminus \bar \Omega\\
   u(x) &= f(x), & x \in \partial \Omega,\\
   \left\| \left\langle \nabla u(x), \frac{x}{\| x \|} 
     \right\rangle - \iota \kappa u(x)\right\| &\in 
     \mathcal O \left(\frac{1}{\lVert x \rVert^2} \right)
     & \lVert x \rVert \to \infty.
\end{align*}
Here $\kappa\in\bbbr_{>0}$ denotes the wave number.
It is well known that if $\kappa$ is close to an eigenvalue
of the interior Neumann problem, the standard 
Dirichlet-to-Neumann mapping \eqref{eqn:varform_laplace} is not 
applicable due to the single layer potential operator not being
coercive anymore.
Well-known proposals to fix this problem are the
Brakhage-Werner approach \cite{BHWP65} and the Burton-Miller
approach \cite{BAJMGF1971}.
We choose the former and reformulate the problem as
\begin{equation*}
 \int_{\partial \Omega} \frac{\partial g}{\partial n(y)}(x,y) w(y) \mathrm d y
   - \iota \eta \int_{\partial \Omega} g(x,y) w(y) \mathrm d y
   = f(x)
\end{equation*}
for all $x\in\bbbr^3\setminus\bar\Omega$,
which has to be solved for the unknown density $w$.
To ensure unique solvability, we have to choose $\eta > 0$.
Applying trace operators yields a boundary integral equation
which we can solve with the same techniques as 
\eqref{eqn:varform_laplace}.

Again we define single and double layer potential operators
$V$ and $K$ using the fundamental solution
\begin{equation*}
  g(x,y) :=  \frac{\exp(\iota \kappa \| x - y \|}
                  {4 \pi \| x - y \|}
\end{equation*}
of the Helmholtz equation.
Now we want to demonstrate the computing power of the two GPGPUs 
combined against all four CPU cores.
In the same way as we did for Laplace's equation, we have chosen the
geometry to be the unit sphere and computed the solution up to discretization
error for different resolutions of the sphere, but with a constant
wave-number $\kappa = 3$.
As Dirichlet data we choose the function
\begin{equation*}
 f(x) = g(x, (0.0, 0.0, 0.2)).
\end{equation*}
Since we do not have an analytic solution for $w$ at our disposal, 
we evaluate the solution $u$ of the Helmholtz equation by the
Brakhage-Werner formulation and compare it to the analytical solution
of the problem in the exterior of the domain, given by the function 
$f$ itself.
The results can be seen in Table~\ref{tab:helmholtz}.

\begin{table*}
\centering
 \caption{Runtimes for quad-core CPU and GPGPU algorithms to setup the
      SLP matrix $V$ the and DLP matrix $K$ for the Helmholtz equation.
      List size is kept constant at 8 MB and 2 worker threads are
      employed for the GPGPUs.} {
  \begin{tabular}{r|r|r|r|r}
  DoFs & CPU SLP & GTX 580 + 680 SLP & CPU DLP & GTX 580 + 680 DLP\\
  \hline
  2048 & 2.3 s & 0.1 s & 2.7 s & 0.1 s\\
  8192 & 9.6 s & 0.3 s & 11.8 s & 0.4 s\\
  32768 & 42.7 s & 1.3 s & 55.2 s & 1.8 s\\
  131072 & 205.8 s & 6.7 s & 263.4 s & 8.9 s\\
  524288 & 843.6 s & 27.9 s & 1050.1 s & 36.9 s
  \end{tabular}}
\label{tab:helmholtz}
\end{table*}


\section{Conclusions}

Matrices arising from the Galerkin discretization of boundary 
integral equations are densely populated and therefore very
compute-intensive.
We have proposed an implementation of the Green cross approximation
algorithm on GPGPUs and have shown that this approach is very fast compared
to pure CPU code, allowing us to set up very large problems quickly.

We have also shown that both the GCA method and our implementation
on the GPGPU work not only for the Laplace equation, but also for
the significantly more challenging Helmholtz equation.

Our implementation is not limited to graphics cards, but can utilize
general computing hardware as long as it supports the OpenCL standard.
The option to use of more that just one accelerator card makes this
approach very interesting for high performance computing applications
because it might be scaled to several hundreds of GPGPUs with some
modifications.

Until now, the setup phase of the boundary integral matrices has
dominated the computing time of boundary element methods.
Using our new CPU/GPGPU approach, the time required for solving the
linear system has become the limiting factor.
This challenge could be approached, e.g., by parallelizing
$\mathcal{H}$- and $\mathcal{H}^2$-matrix preconditioners
\cite{GRHA02,BORE14,KR13}.


\bibliographystyle{plain}
\bibliography{scicomp}

\end{document}